\documentclass[aps, prl, twocolumn, floatfix, a4paper]{revtex4-1}  
\usepackage{graphicx}
\usepackage{dcolumn}
\usepackage{array}
\usepackage[usenames]{color}
\usepackage{amsmath}
\usepackage{amssymb}
\usepackage{ulem}
\usepackage[paperwidth=210mm,paperheight=297mm,centering,hmargin=1.75cm,vmargin=2cm]{geometry}
\begin{document}
\newlength{\imgsz}
\setlength{\imgsz}{0.9\columnwidth}

\title{Efficient first-principles calculation of the quantum kinetic energy\\ and  
momentum distribution of nuclei}

\author{Michele Ceriotti}
\email{michele.ceriotti@chem.ox.ac.uk}
\affiliation{Physical and Theoretical Chemistry Laboratory, 
University of Oxford, South Parks Road, Oxford OX1 3QZ, UK}

\author{David E. Manolopoulos}
\affiliation{Physical and Theoretical Chemistry Laboratory, 
University of Oxford, South Parks Road, Oxford OX1 3QZ, UK}

\begin{abstract}
Light nuclei at room temperature and below exhibit a kinetic energy which
significantly deviates from the predictions of classical statistical
mechanics.  This quantum kinetic energy is responsible for a wide variety of isotope 
effects of interest in fields ranging from chemistry to climatology. It also furnishes
the second moment of the nuclear momentum distribution, which contains 
subtle information about the chemical environment and has recently become
accessible to deep inelastic neutron scattering experiments.
Here we show how, by combining imaginary time path integral 
dynamics with a carefully designed generalized Langevin equation, it is
possible to dramatically reduce the expense of computing
the quantum kinetic energy. We also introduce a transient anisotropic Gaussian
approximation to the nuclear momentum distribution which can be calculated
with negligible additional effort. As an example, we evaluate the structural properties, 
the quantum kinetic energy, and the nuclear momentum distribution
for a first-principles simulation of liquid water.
\end{abstract}
\maketitle

\setlength{\textfloatsep}{1em plus 0.25em minus 0.25em}

One of the most commonly adopted approximations in atomistic simulations of
materials and chemical compounds is the assumption that atomic nuclei behave 
as classical particles. Unfortunately, whenever
hydrogen or other light nuclei are present and the simulation is performed 
at or below room temperature, significant deviations from classical 
behaviour are to be expected. Moreover, many interesting physical phenomena and 
experimental observables depend directly on the quantum nature of the nuclear motion.
This implies that an explicit treatment of nuclear quantum effects (NQE) may
not only be desirable to improve the accuracy of the simulation; it can 
even be essential for comparison with experiments.
In particular, the quantum kinetic energy (QKE) underlies {\em all} changes in
relative free energy associated with isotopic substitution, 
which are responsible for phenomena as diverse as
isotope effects in enzyme catalysis and the fractionation of isotopes between
different compounds and phases. The accurate evaluation of NQE therefore
has implications for disciplines ranging from chemistry, biochemistry, and 
materials science to geology, archaeometry, and climatology
 \cite{west61chrev,*cha+89science,*flan-oate91arms,*frie-onei77book,*savi-epst70gca,*gat96areps}.

The state-of-the-art method for computing the QKE involves the use
of path integral molecular dynamics (PIMD) \cite{parr-rahm84jcp}. This allows one to systematically 
converge NQE but introduces an often prohibitive overhead with respect
to a simulation with classical nuclei, since one has to compute the energy
of many replicas of the physical system. Recently it has been suggested
that a generalized Langevin equation (GLE) can be used to model
NQE less expensively \cite{ceri+09prl2,buyu+08pre,*damm+09prl}, and that systematic convergence
can be obtained by combining path integrals (PI) with an appropriate
GLE \cite{ceri+11jcp}. 
In the present Letter we generalize this PI+GLE approach in such a way
that different quantum mechanical observables, including the QKE, 
can be obtained at a fraction of the cost of a full PI calculation. 
We also introduce a convenient ``transient anisotropic Gaussian'' (TAG) approximation
to the nuclear momentum distribution $n(p)$. This significantly simplifies the task of comparing
first-principles atomistic modeling with deep inelastic neutron scattering (DINS) 
 \cite{andre+05advp,*pant+08prl,*andr+11cpl,*flam+12jcp} -- a promising experimental 
 technique which directly probes the quantum nature of light atoms.


Path integral molecular dynamics is based on the isomorphism between
the quantum mechanical partition function of a system of $N$ distinguishable
particles and the classical partition function of a so-called ring polymer \cite{chan-woly81jcp}, 
described by the Hamiltonian
\begin{equation}
H_P\left(\mathbf{p},\mathbf{q}\right) = \sum_{i=0}^{P-1} \frac{1}{2}\mathbf{p}_i^2 + 
V\left(\mathbf{q}_i\right) + \frac{1}{2}\omega_P^2 \left(\mathbf{q}_i-\mathbf{q}_{i+1}\right)^2.
\label{eq:pi-hamiltonian}
\end{equation}
Here $P$ is the number of replicas (beads) composing the path, $\mathbf{q}_i$ and $\mathbf{p}_i$ 
are $3N$-dimensional vectors describing the mass-scaled positions and momenta of the 
particles in the $i$-th replica, and $V(\mathbf{q}_i)$ is the physical potential acting on replica $i$. 
For a simulation at the physical
temperature $1/k_B\beta$, the harmonic interaction between neighbouring beads
is characterized by the frequency $\omega_P=P/\beta\hbar$, and the 
ring polymer Hamiltonian~\eqref{eq:pi-hamiltonian} must be sampled 
at $P$ times the physical temperature.

As the number of replicas is increased, the equilibrium properties computed
from the simulation converge to the correct quantum mechanical expectation values. 
In particular, one can compute the average potential
\begin{equation}
\left<V\right> = \frac{1}{P}\sum_{i=0}^{P-1}\left<V(\mathbf{q}_i)\right>
\label{eq:potential}
\end{equation}
and the quantum kinetic energy (using the centroid virial estimator \cite{herm-bern82jcp,cepe95rmp})
\begin{equation}
\left<T\right> = \frac{3N}{2\beta} + \frac{1}{2P}\sum_{i=0}^{P-1}\left<\left(\mathbf{q}_i-\bar{\mathbf{q}}\right)\cdot \nabla V\left(\mathbf{q}_i\right)\right>,
\label{eq:kinetic}
\end{equation}
where $\bar{\mathbf{q}}=\sum_i \mathbf{q}_i/P$ is the centroid of the ring polymer.
The convergence of the averages~\eqref{eq:potential} and~\eqref{eq:kinetic} 
to the quantum expectation values for a system whose fastest normal mode has
frequency $\omega_{\rm max}$ requires a number of replicas of the order
of $\beta\hbar\omega_{\rm max}$, making the PI approach very demanding when simulations
are performed at low temperature, or  in the presence of stiff vibrations. 

It has recently been shown that this overhead can be removed by using correlated noise
to construct a tailored, non-equilibrium Langevin dynamics that mimics NQE
using a single replica \cite{ceri+09prl2}. This approach can be made exact in the 
harmonic limit, without requiring explicit knowledge of the Hessian, 
but it is inherently approximate in any real system containing anharmonicities. 
To address this shortcoming, one can combine correlated noise with PIMD \cite{ceri+11jcp}. 
In this PI+GLE approach, each replica is subject to an independent instance of the same GLE
thermostat, which is designed in such a way that the expectation value $\left<V\right>$ 
is exact for any number of replicas in the harmonic limit. Convergence is approached systematically, 
such that the computational effort required to obtain structural properties is reduced by a factor of 
4 or more relative to PIMD \cite{ceri+11jcp}.

Note, however, that while the average potential~\eqref{eq:potential} only depends on the 
positions of individual beads, the kinetic energy~\eqref{eq:kinetic}
involves their centroid $\bar{\bf q}$, and 
therefore depends on cross-correlations between the coordinates of different
beads. The PI+GLE method described in Ref.~\cite{ceri+11jcp} does not address 
the convergence of these correlations, and therefore $\left<T\right>$ is not guaranteed 
to be exact in the harmonic limit. As a result, the kinetic energy converges more slowly than 
the potential energy, requiring about twice the number of replicas to reach the same level of accuracy.
Fortunately, it is relatively simple to improve PI+GLE to also manipulate bead-bead 
correlations. It is in fact sufficient to apply different GLE thermostats
to the different normal modes of the ring polymer. The desired correlations
can then be enforced in the harmonic limit, by separately tuning the 
parameters of the various thermostats.

This idea is readily applied to the problem of designing a ``PIGLET'' method, 
which converges $\left<T\right>$ just as rapidly as $\left<V\right>$
\footnote{Note that this approach could easily be extended to manipulate the 
bead-bead correlations in a PI+GLE simulation even further, so as to accelerate
the convergence of other PI estimators -- for example those used to compute 
imaginary-time correlation functions. }. 
Evaluating~\eqref{eq:potential} and~\eqref{eq:kinetic} for a one-dimensional harmonic oscillator of 
frequency~$\omega$ yields
\begin{equation}
\begin{split}
\left<T\right>=&\frac{1}{2\beta} + \frac{1}{2P}\omega^2 \sum_{i=0}^{P-1}\left<q_i^2\right> - \frac{1}{2}\omega^2\left<\bar{q}^2\right>=\\
=&\left<V\right>+\frac{1}{2\beta} - \frac{1}{2}\omega^2\left<\bar{q}^2\right>.
\end{split}
\end{equation}
Therefore, in order to enforce the quantum conditions
\begin{equation}
\left<V\right>=\left<T\right>=\frac{\hbar\omega}{4}\coth {\beta\hbar\omega\over 2},
\end{equation}
one sees that the centroid must be distributed classically, leaving the fluctuations of 
the individual beads to impose the quantum statistics. The simplest way to ensure this
is to apply a classical thermostat to the centroid coordinate, 
and the same non-equilibrium GLE to all the internal modes of the ring polymer.
This second thermostat must be designed to enforce the correct quantum fluctuations
on each normal mode, in accordance with (5). The desired frequency-dependence of the
thermostat temperature can be obtained by solving a functional equation analogous to Eq.~(16) 
in Ref.~\cite{ceri+11jcp}. Parameters for such a GLE thermostat have been generated using
the fitting procedure described in Ref.~\cite{ceri+10jctc}, and may be found in the
supplementary material \cite{EPAPS} or downloaded from~\cite{gle4md}.

\begin{figure}
\includegraphics[width=\imgsz]{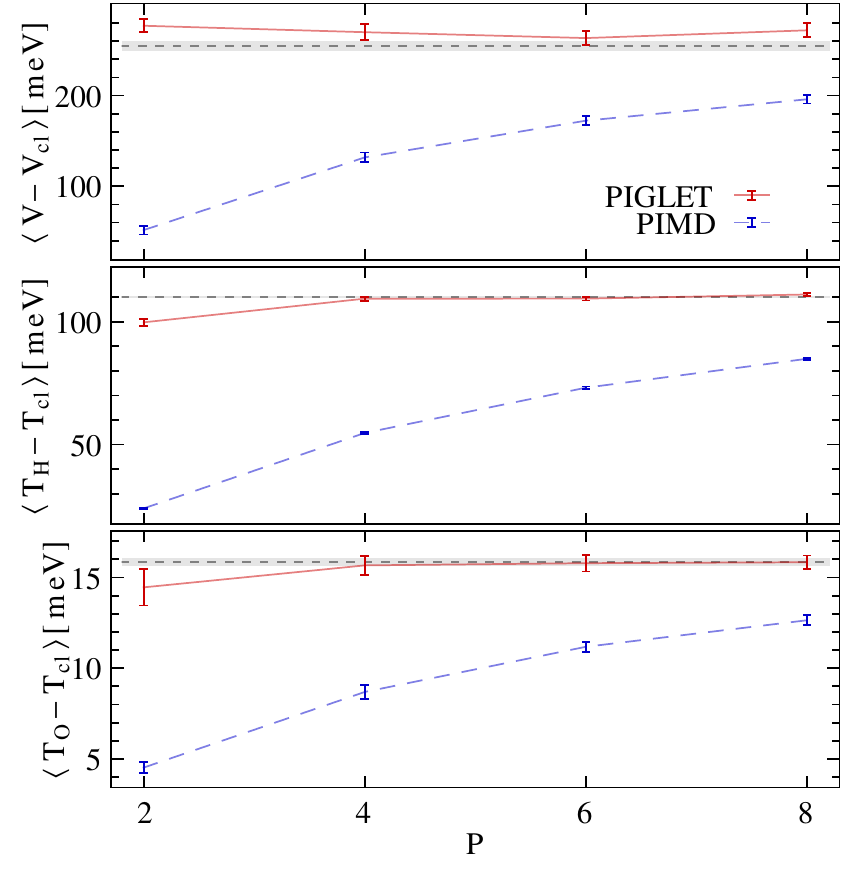}
\caption{\label{fig:piglet-water} 
Per-molecule quantum contribution to the potential energy and per-atom quantum kinetic 
energy plotted as a function of the number of PI replicas. The convergence of 
conventional PIMD (blue points) is compared with the present method (red points). 
Data points are reported with $2\sigma$ error bars, which have been inferred from 
longer simulations using an empirical water model \cite{EPAPS}. In each panel, the dashed line
corresponds to the average computed from a $32$-bead PIMD reference calculation, and the shaded
area to a $2\sigma$ confidence interval.}
\end{figure}

\begin{figure}
\includegraphics[width=\imgsz]{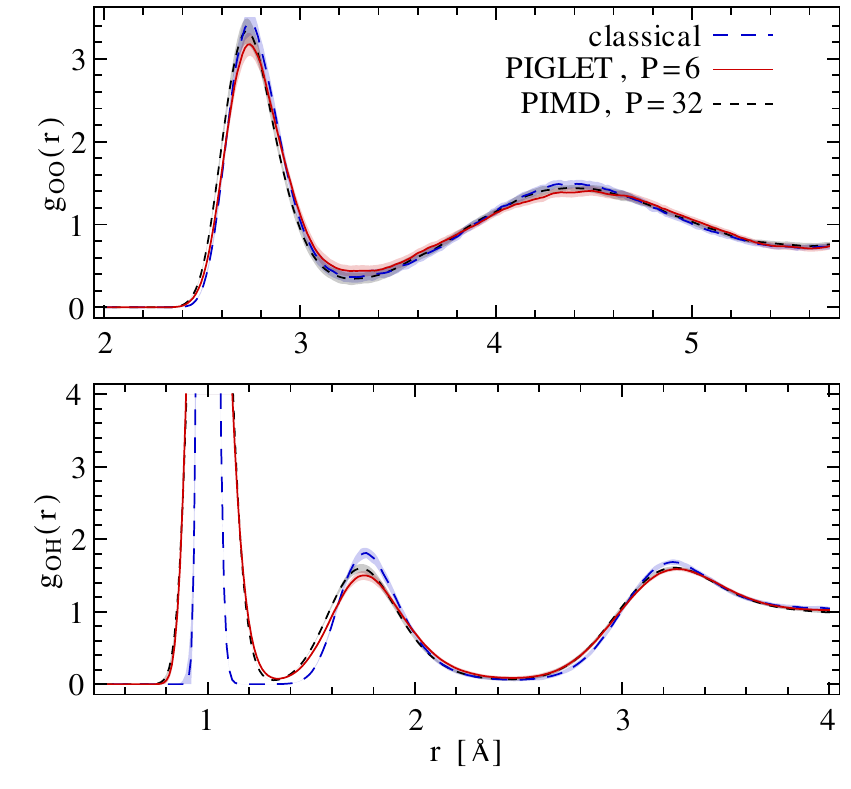}
\caption{\label{fig:piglet-gdr} Radial distribution functions for the O--O (upper panel) and 
O--H (lower panel) pairs. Results from classical MD are reported in blue, results from a 
PIGLET simulation with 6 beads are reported in red. A reference calculation with conventional 
PIMD using 32 beads is reported as a black, dashed line. Statistical confidence 
intervals are represented as shaded areas and have been inferred
from longer simulations using an empirical water model \cite{EPAPS}.
}
\end{figure}

To illustrate the potential of this PIGLET method, we have performed a simulation
of liquid water modelled using density functional theory,
as implemented in the CP2K code \cite{vand-krac05cpc,*goed+96prb}. We used the BLYP exchange-correlation functional \cite{beck88pra,*lee+88prb}, 
and the same well-established simulation details as described in Ref.~\cite{hass+11pnas}.
We performed simulations of a box of 64 water molecules at a temperature of $300$~K and the experimental density.
We used a time step of $0.5$~fs, and each simulation was $15$~ps long with the first $3$~ps 
discarded for equilibration. 32-bead PIMD simulations required a shorter time step of $0.25$~fs, and 
were only $7.5$~ps long.

Figure~\ref{fig:piglet-water} demonstrates the fast convergence of $\left<V\right>$ and $\left<T\right>$, 
both of which reach a level of accuracy within the statistical error of a standard $32$-bead PIMD 
simulation using just 6 beads.  
This dramatic reduction in the computational effort required to compute the QKE would 
for instance make it possible to evaluate isotope fractionation
between the liquid and the gas phases of water by first-principles simulations.
In this context, the delicate balance between the competing quantum effects 
which characterise the behaviour of water
\cite{habe+09jcp,li+11pnas,mark-bern12pnas} would provide a sensitive benchmark
for the relative merits of different exchange-correlation functionals.
Furthermore, the increased efficiency in computing the QKE does not come at the expense of the 
accuracy of structural properties: in Figure~\ref{fig:piglet-gdr} we demonstrate that the O-H and O-O  
radial distribution functions computed with PIGLET using 6 beads also agree
with those from a 32-bead PIMD simulation to within the statistical accuracy.

\begin{figure}
\includegraphics[width=\imgsz]{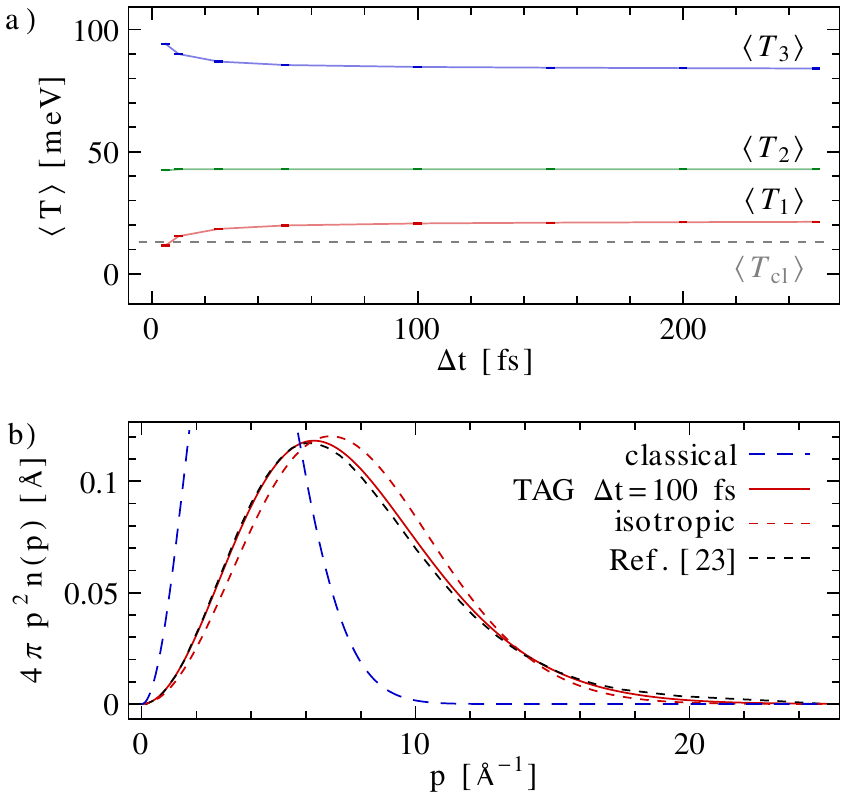}
\caption{\label{fig:piglet-np} 
Panel (a) shows the principal values of the TAG kinetic energy tensor
as a function of the moving average window width $\Delta t$, for a 6-bead PIGLET simulation
of liquid water at $300$~K. 
Panel (b) compares the classical $n(p)$ for a proton at $300$~K (blue curve) with the TAG $n(p)$
obtained using $\Delta t=100$~fs. Other values of $\Delta t$ in the range $100<\Delta t<500$~fs 
give the same result to graphical accuracy.
Note the nearly-perfect agreement with the open-path $n(p)$ from Ref.~\cite{morr-car08prl} (black dashed line),
which is in turn in very good agreement with experimental data~\cite{reit+04bjp,*giul+11prl,morr-car08prl}.
The TAG gives a significant improvement over an isotropic Gaussian approximation (red dashed line).}
\end{figure}

The quantum kinetic energy is also directly related to the second moment of the  
distribution of particle momentum $n(p)$, which can be measured in DINS experiments.
The nuclear momentum distribution, however, contains more information
about the chemical environment of the nucleus than the kinetic energy alone.
In order to access $n(p)$ computationally, the state-of-the-art technique involves
performing an open path or displaced path PIMD simulation, both of which are even 
more complex and computationally demanding
than conventional PIMD \cite{lin+10prl}. 

Fortunately, it turns out that $n(p)$ can often be modelled very accurately as resulting 
from the spherical average of an anisotropic three-dimensional Gaussian distribution 
\cite{pant+08prl,lin+11prb}. For an individual atom in a rigid environment, one could imagine 
constructing
such a distribution by computing the principal components of its kinetic energy tensor, the virial
estimator for which is
\begin{equation}
T_{\alpha\beta}=\frac{\delta_{\alpha\beta}}{2\beta} + \frac{1}{4P} 
\sum_i \left[\left(q_{i\alpha}-\bar{q}_{\alpha}\right) \frac{\partial V}{\partial q_{i\beta}} 
+ \left(q_{i\beta}-\bar{q}_{\beta}\right) \frac{\partial V}{\partial q_{i\alpha}}\right].
\label{eq:kinetic-tensor}
\end{equation}
Here $q_{i\alpha}$ indicates Cartesian component $\alpha$ of the position of the $i$-th replica
of the atom in the ring polymer.

If one were to compute the average of \eqref{eq:kinetic-tensor} for an atom in liquid water, however,
an isotropic tensor would be obtained, because of the ever-changing orientation of the water molecules.
On the other hand, the fact that the experimental $n(p)$ agrees with an anisotropic Gaussian model 
suggests that \emph{instantaneously} each atom has a quasi-Gaussian momentum distribution, 
albeit in a dynamic, transient frame of reference. This idea suggests a very simple approximation to
$n(p)$ based on a moving average of the kinetic energy tensor:
\begin{equation}
T_{\alpha\beta}(t;\Delta t)=\frac{1}{\Delta t} 
\int_{t-\Delta t}^{t+\Delta t} T_{\alpha\beta}(t') \left(1-\frac{\left|t-t'\right|}{\Delta t}\right) \mathrm{d}t'.
\label{kinetic-moving}
\end{equation}
Suppose that we evaluate the eigenvalues $T_\gamma\left(t;\Delta t\right)$ of this tensor (sorted in 
increasing order) at each instant $t$, and compute their averages
$\left<T_\gamma(\Delta t)\right>$ along the trajectory. Then these averages can be
used to define a \lq\lq transient anisotropic Gaussian" (TAG) approximation
to $n(p)$, with $\left<p_{\gamma}^2\right>=2\left<T_{\gamma}(\Delta t)\right>$ for $\gamma=1,2,3$ \footnote{Note that $\sum_\gamma\left<T_\gamma(\Delta t)\right>=\left<T\right>$
by construction, so that the $n(p)$ computed within this TAG approximation will automatically
be consistent with the total QKE.}.

As shown in Figure~\ref{fig:piglet-np}(a), all three $\left<T_\gamma(\Delta t)\right>$ 
converge to plateau values after a characteristic time interval $\Delta t$ -- the time
needed to establish the principal axes of the TAG. 
On a much longer time scale, if the atom in question experiences many different transient
environments, one would expect all three $\left<T_\gamma(\Delta t)\right>$ to drift, 
eventually approaching $\left<T\right>/3$ in an isotropic system such as a liquid.
In Figure~\ref{fig:piglet-np}(b) we show that the spherical average of the
TAG generated with the plateau values of $\left<T_\gamma(\Delta t)\right>$ 
almost perfectly matches the results obtained with heroic effort by Morrone and Car
\cite{morr-car08prl} 
using an open path integral simulation, and differs significantly from the $n(p)$ constructed from an
isotropic Gaussian with $\left<p^2\right>=2\left<T\right>$. 

The TAG approximation shares some similarities with the Gaussian approximation to the open path 
distribution introduced in Ref.~\cite{lin+11prb}, and with the idea of computing the components
of the kinetic energy along a set of orthogonal vectors anchored to the molecular axes of individual
water molecules \cite{mark-bern12pnas}\footnote{If the $n(p)$ of individual particles were to 
exhibit significant non-Gaussian behaviour, neither of these techniques nor the TAG approach 
would be a substitute for a much more demanding open path simulation.}. Some of the advantages of the 
TAG approach are that: (1) it is not necessary to open any paths, so there is no additional sampling 
overhead compared with 
a conventional PIMD simulation; (2) since all paths remain closed, one can increase the statistical 
efficiency by averaging the computed $n(p)$ over all equivalent atoms; (3) it is not restricted to crystals  
or to systems with identifiable molecular axes -- the local frame of reference of each atom does not have to be specified in advance but is determined automatically; and (4) it comes with an internal sanity check, 
since one can verify the existence of a transient local environment by checking
that all three $\left<T_\gamma(\Delta t)\right>$ reach well-defined plateaus.

\begin{table} 
\begin{ruledtabular}
\begin{tabular}{ c d c c c c }
  (meV)     & \multicolumn{1}{c}{$\left<T\right>$} & (Exp; Temp.)    &  $\left<T_1\right>$  & $\left<T_2\right>$  &$\left<T_3\right>$ \\
\hline
H (H$_2$O)  &   148.2 &  (145; 296K)          &  20.6                & 42.8                & 84.7              \\
D (D$_2$O)  &   110.1 &  (106$\pm$5; 292K)          &  16.4                & 31.4                & 62.4              \\
O (H$_2$O)  &    54.6 &            &  14.4                & 18.3                & 21.8              \\
O (D$_2$O)  &    58.1 &            &  14.5                & 19.6                & 24.0              \\
\end{tabular}
\end{ruledtabular}
\caption{\label{tab:qke} Per-atom QKE and its TAG components in simulations
of light and heavy water at $300$~K, computed using
$\Delta t=100$~fs.  
NQE were described by the PIGLET method with $6$ beads. All the statistical
errors are  smaller than $0.1$~meV. Where available, experimental results at a similar
temperature are reported in parentheses \cite{reit+04bjp,*giul+11prl}. }
\end{table}

The TAG $n(p)$ is also sufficiently accurate to reveal some rather subtle effects, which 
may be detectable in DINS experiments that are currently underway \cite{senesi}.
In a comparative study of light and heavy water, one finds that while the
D nuclei have a lower QKE in D$_2$O than the H nuclei in H$_2$O, the oxygen nuclei actually
have a {\em higher} QKE in D$_2$O than in H$_2$O. This observation can be
rationalised in terms of a reduced mass effect, whereby the O nucleus in heavy water shares a higher 
fraction of the intra-molecular zero-point energy than the oxygen in light water. 
As shown in Table~\ref{tab:qke}, the effect is clearly captured by the TAG approximation. 
One of the components of the O momentum distribution, orthogonal to 
the plane of the molecule, is almost unaffected by the isotope substitution, while the 
components which contribute to the intra-molecular vibrations possess a higher 
QKE in D$_2$O than in H$_2$O. 

The methods we have introduced in this Letter clearly simplify the task
of computing properties related the quantum nature of light nuclei, potentially
enabling comparison with experimental observations in more complex cases than have
hitherto been possible. The resulting interplay between theory 
and experiment will be essential to unravel the competing
quantum effects that govern the behaviour of water and
other hydrogen-bonded systems \cite{li+11pnas}.

This work was supported by the Royal Society, the Swiss National Science Foundation, 
the Wolfson Foundation, and the EU Marie Curie IEF n. PIEF-GA-2010-272402. 
Computer time from the University of Lugano is gratefully acknowledged.

\end{document}